\documentclass[english, citeautoscript, physrev, prb, superscriptaddress, twocolumn]{revtex4-2}
\usepackage[T1]{fontenc}
\usepackage[latin9]{inputenc}
\usepackage{color}
\usepackage{verbatim}
\usepackage{algorithm2e}
\usepackage{amsmath}
\usepackage{amssymb}
\usepackage{graphicx}
\usepackage{caption}
\makeatletter
\@ifundefined{showcaptionsetup}{}{%
 \PassOptionsToPackage{caption=false}{subfig}}
\usepackage{subfig}
\usepackage[normalem]{ulem}
\makeatother

\usepackage{hyperref}
\hypersetup{
    colorlinks=true,
    linkcolor=blue,
    filecolor=magenta,      
    urlcolor=blue,
    citecolor=blue,
}

\usepackage{babel}
\begin{document}
\title{Dynamical tuning of the chemical potential to achieve a target particle number in grand canonical Monte Carlo simulations}
\author{Cole Miles}
\email{cmm572@cornell.edu}
\affiliation{Department of Physics, Cornell University, Ithaca, New York 14850, USA}
\author{Benjamin Cohen-Stead}
\affiliation{Department of Physics, University of California, Davis, California 95616, USA}
\author{Owen Bradley}
\affiliation{Department of Physics, University of California, Davis, California 95616, USA}
\author{Steven Johnston}
\affiliation{Department of Physics and Astronomy, The University of Tennessee, Knoxville, Tennessee 37996, USA}
\affiliation{Institute for Advanced Materials and Manufacturing, University of Tennessee, Knoxville, Tennessee 37996, USA\looseness=-1}
\author{Richard Scalettar}
\affiliation{Department of Physics, University of California, Davis, California 95616, USA}
\author{Kipton Barros}
\email{kbarros@lanl.gov}
\affiliation{Theoretical Division and CNLS, Los Alamos National Laboratory, Los
Alamos, New Mexico 87545, USA}
\begin{abstract}
We present a method to facilitate Monte Carlo simulations in the grand canonical ensemble given a target mean particle number.
The method imposes a fictitious dynamics on the chemical potential, to be run concurrently with the Monte Carlo sampling of the physical system.
Corrections to the chemical potential are
made according to time-averaged estimates of the mean and variance of the particle number, with the latter being proportional to thermodynamic compressibility.
We perform a variety of tests, and in all cases find rapid convergence of the chemical potential---inexactness
of the tuning algorithm contributes only a minor part of the total measurement error for realistic simulations.
\end{abstract}
\maketitle
\global\long\def\d{\mathrm{d}}%
\global\long\def\tr{\mathrm{\mathrm{tr}}\,}%

\section{Introduction}

A fundamental attribute of statistical mechanics is the equivalence
of thermodynamic ensembles in the limit of large system size. In particular, the \emph{canonical} ensemble, with fixed particle number, should
be equivalent to a \emph{grand} canonical ensemble in which
the chemical potential $\mu$ is suitably selected to fix the average particle number. However, there
may be practical reasons to prefer working in the grand canonical ensemble, particularly in the context of Monte Carlo (MC)
simulations. In classical
MC simulations, for example, moves that modify the particle number can be useful for
reducing decorrelation times or for studying coexistence between
phases~\citep{Yao82, Escobedo96, Kowalczyk05, Clark06, Eslami07}.
Similarly, the starting point of many~\cite{Blankenbecler81,Batrouni85,Scalettar87, White89,kawashima94,zhang99,he19} (but not all~\cite{hirsch82, sandvik97, zhang95, zhang97}) finite temperature quantum Monte Carlo (QMC) simulations
is the grand canonical partition function  $Z=\mathrm{Tr}\,\exp[-\beta(H+\mu N)]$, where $\beta$ is the inverse temperature, $H$ is the Hamiltonian,
and $N$ is the number operator. The trace above runs over \emph{all} quantum wavefunctions, not just those constrained to a fixed particle number. 

In MC and QMC simulations, we often wish to specify the average particle number $\left\langle N\right\rangle$ directly, e.g., to fix the electron filling fraction. Determining the $\mu$ value which satisfies this condition has traditionally required a tedious manual search, with additional
searches necessary after every update to the model parameters.
Here we present a method to efficiently converge the chemical potential
$\mu$ to a solution value that produces the desired mean particle number within the same MC/QMC simulation where measurements are performed.
We take the chemical potential $\mu_{t}$ to be continually evolving
in sampling time $t$. Corrections to $\mu_{t}$ are performed whenever new
measurements $N_{t}$ of $\left\langle N\right\rangle $ are collected. 

A central challenge is that, in certain cases, it can be difficult to collect good statistical samples for systems with long autocorrelation times. That is, modifications to $\mu_{t}$ may not
fully impact the samples $N_{t}$ until quite some time later. We
address this problem by employing increasingly long-time averages, which
incorporate a fixed fraction of \emph{the entire history of MC data.}

Roughly speaking, our proposed strategy for tuning $\mu_{t}$ is as
follows: Given continually improving approximate measurements of the
particle number $\left\langle N\right\rangle $ and compressibility
$\kappa=\mathrm{d}\left\langle N\right\rangle /\mathrm{d}\mu$, we update the value of the chemical potential under the assumption of linear response. Our method shares some conceptual similarities with proportional--integral--derivative (PID) controllers, which
have previously been applied to $\mu$-tuning~\citep{Speidel06,Kleiner19}.
A disadvantage of PID controllers is that they introduce several
parameters that must be carefully selected for each new problem. In
contrast, the method we introduce here is simple and works robustly
across a wide range of problems using a single default set of algorithm parameters.

We benchmark the new method on two problems: (1) Tuning the applied field in the classical two-dimensional
ferromagnetic Ising model to achieve a target magnetization, and (2) tuning the chemical potential in simulations of the quantum Holstein model to achieve a target electron filling fraction. For the Ising model, simulation temperatures approaching $T_{c}$ give rise to long autocorrelation times, which can make it difficult to achieve good statistical sampling. For QMC simulations of the Holstein model, phonons mediate
an effective attractive electron-electron interaction, which, in turn, gives rise to challenging metastability effects.

The benchmarks indicate that accurate measurements of a system with a specified mean particle number can be acquired from a {\em single} MC simulation run, with $\mu$-tuning enabled throughout. 
This works because $\mu_t$ mostly converges within the burn-in period of the MC simulation. After burn-in there remains a small (and steadily decreasing) error in $\mu_t$, but it does not seem to contribute significantly to the overall statistical error in measurements.

\section{Method\label{sec:Method}}

We present a method to tune any thermodynamic field 
according to its conjugate observable. For concreteness, consider the task of tuning the chemical potential $\mu$
to produce a target mean particle number, $\left\langle N\right\rangle = N^{\ast}$. The same method could be applied to a magnetic system, in which case one would replace $\mu$ with the applied magnetic field and $\left\langle N\right\rangle $ with the total magnetization (an example is considered in Sec.~\ref{subsec:Ising}).

At fixed $\mu$, one can estimate the mean particle number $\langle N\rangle$
using MC sampling. The compressibility
$\kappa$ will play an essential role in our tuning scheme. A fundamental result from thermodynamics states
\begin{equation}
\kappa=\frac{\d\langle N\rangle}{\d\mu}=\beta\,\mathrm{Var}[N],\label{eq:kappa}
\end{equation}
where $\mathrm{Var}[N]=\langle N^{2}\rangle-\langle N\rangle^{2}$.
Thus, $\kappa$ can be estimated using the observed variance of $N$.

\subsection{Prior work with iterated simulation\label{subsec:windowed}}

Previous work proposed the following $\mu$-tuning strategy~\citep{Speidel06}:
At fixed $\mu$, run MC over some time window to collect statistical
estimates $\overline{N}$ and $\overline{\kappa}$ of the mean particle number and compressibility.
To find the chemical potential that will approximately achieve a target
particle number $N^{\ast}$, solve for the $\mu^\prime$ value that satisfies
$\overline{\kappa} = (N^{\ast}-\overline{N})/(\mu^\prime-\mu)$. Assign
$\mu \leftarrow \mu^\prime$ and repeat.

A practical challenge with this iterated update scheme is that it
may be difficult to acquire sufficiently good estimates $\overline{N}$
and $\overline{\kappa}$. It is hard to know \emph{a priori} how much
sampling time should be devoted to any particular $\mu$ value. Statistical estimates
of $\kappa$, associated with fluctuations in $N$, are particularly
error prone. Also, if many iterative updates to $\mu$ are required,
it would seem advantageous to incorporate information
from \emph{all} previous MC runs, not just from the most recent
sampling window.

\subsection{Dynamical \texorpdfstring{$\boldsymbol{\mu}$}{μ}-tuning}
In contrast to the iterated simulation scheme,
here we explore an approach where $\mu_{t}$ evolves dynamically in
the context of a single simulation. At each iteration $t$, a
MC update step or sweep is performed using the instantaneous value $\mu_{t}$
of the chemical potential. Next, the chemical potential is updated
using the rule
\begin{equation}
\mu_{t+1}=\overline{\mu}_{t}+(N^{\ast}-\overline{N}_{t})/\overline{\kappa}_{t}.\label{eq:tuner}
\end{equation}
We use the notation $\overline{\left(\cdot\right)}_{t}$ to signify an appropriate
time average over a subset of the sampled data up to time $t$. Note that \emph{the
effective window size is continually increasing} \emph{with sampling
time}. Many types of time-averaging are possible; for simplicity,
we average over the most recent half of all collected data, weighting
each sample equally. Averages of $\mu$ and $N$ up to time $t$ are
defined as 
\begin{align}
\overline{\mu}_{t} & =\frac{1}{L_{t}}\sum_{t^\prime=\left\lceil t/2\right\rceil }^{t}\mu_{t^\prime}\label{eq:mu_bar}\\
\overline{N}_{t} & =\frac{1}{L_{t}}\sum_{t^\prime=\left\lceil t/2\right\rceil }^{t}N_{t^\prime}.\label{eq:N_bar}
\end{align}
The ceiling function $\left\lceil \cdot \right\rceil $ rounds up to the nearest integer
and $L_{t}\equiv t-\left\lceil t/2\right\rceil +1$ is the number of samples
in the average. We select this form for the running-time averages partly for simplicity, and partly because it allows updates in constant time, as described in Appendix~\ref{sec:welford}.

An important aspect of Eq.~(\ref{eq:tuner}) is that it defines $\mu_{t+1}$ as a correction to the time-averaged chemical potential
$\overline{\mu}_{t}$, and \emph{not} as a correction to the previous instantaneous chemical potential $\mu_{t}$. In this way, the estimator $\mu_{t}$ captures important information from the entire sampling history, and evolves
on the same time scale as $\overline{N}_{t}$.

\subsubsection{Estimating compressibility}
The success of Eq.~(\ref{eq:tuner}) depends crucially on the definition
of the time-averaged compressibility $\overline{\kappa}_{t}$. Equation~(\ref{eq:kappa})
suggests that we can estimate $\kappa$ using the time-averaged variance,
\begin{equation}
\kappa_{t}^{\mathrm{fluc}}=\beta\,\overline{\mathrm{Var}}_{t}[N].\label{eq:kappa_fluc}
\end{equation}
For classical systems, $\overline{\mathrm{Var}}_{t}[N]$ is defined
as the sample variance for the data $\{N_{t/2},\dots N_{t}\}$. For
quantum systems, a slightly modified definition is given in Eq.~(\ref{eq:quantum_var}).
Generally, the fluctuation-based estimator $\kappa_{t}^{\mathrm{fluc}}$
becomes valid at late times, once $\mu_{t}$ settles to a near constant
value. At early times, however, $\mu_{t}$ is evolving rapidly, and
the estimator $\kappa_{t}^{\mathrm{fluc}}$ is error prone. To ensure
that the update rule of Eq.~(\ref{eq:tuner}) is reasonable at all
times, we impose carefully defined lower and upper bounds for our
final compressibility estimator,

\begin{equation}
\overline{\kappa}_{t}=\max\left[\kappa_{t}^{\mathrm{min}},\min\left(\kappa_{t}^{\mathrm{max}},\kappa_{t}^{\mathrm{fluc}}\right)\right].\label{eq:kappa_eff}
\end{equation}
This ordering of the max and min operators ensures
that $\overline{\kappa}_{t}$ never vanishes.

Imposing the \emph{lower }bound $\kappa_{t}^{\min}$ protects against
the case where early-time fluctuations of $N$ are artificially small.
This could happen in a QMC simulation, for example, if the initial
guess $\mu_{t=0}$ is in one of the system's band gaps ($N_t$ associated with fully occupied bands), leading to a divergence in $1 / \kappa^{\mathrm{fluc}}_{t}$. Since the error in statistical observables
decays like the inverse square root of the number of samples, a reasonable
lower bound on $\overline{\kappa}_{t}$ is given by 
\begin{equation}
\kappa_{t}^{\min}=\frac{\alpha}{\sqrt{t+1}},\label{eq:kappa_min}
\end{equation}
for some appropriately defined $\alpha$. Referring to Eq.~(\ref{eq:kappa}),
we see that $\kappa$ should scale like the system volume $V$ divided
by an intensive energy. For the benchmarks
in Sec.~(\ref{sec:Results}), we will select $\alpha=N_\mathrm{sites}/u_{0}$ where $N_\mathrm{sites}$ is the number of
lattice sites, and $u_{0}$ is a characteristic energy scale. Our tests indicate that algorithm performance is largely insensitive to the precise choice of $\alpha$. By design,
$\kappa_{t}^{\min}$ decays to zero at large times $t$.

Imposing the \emph{upper }bound $\kappa_{t}^{\mathrm{max}}$ protects
against the case where early-time fluctuations of $N$ are artificially
large. Although Eq.~(\ref{eq:kappa_fluc}) is correct in thermodynamic
equilibrium, it produces a very poor estimator in the \emph{out-of-equilibrium}
context of a dynamically evolving $\mu_{t}$. Changes to $\mu_{t}$
will, by design, generate a strong response in $N_{t}$. Drift in
$N_{t}$ will cause a large overestimate of the compressibility, $\kappa_{t}^{\mathrm{fluc}}\sim\overline{\mathrm{Var}}_{t}[N]\sim V^{2}$.
Recall that physical compressibility must scale like system size,
$\kappa\sim V$, which is apparent from the definition $\kappa=\d\langle N\rangle/\d\mu$.
To get a compressibility estimator with the correct scaling, we can
compare the typical variations in $N_{t}$ and $\mu_{t}$.
Specifically, we define the upper bound on $\overline{\kappa}_{t}$
to be
\begin{equation}
\kappa_{t}^{\mathrm{max}}=\sqrt{\frac{\overline{\mathrm{Var}}_{t}[N]}{\overline{\mathrm{Var}}_{t}[\mu]}},\label{eq:kappa_max}
\end{equation}
where $\overline{\mathrm{Var}}_{t}[\mu]$ is the sample variance for
$\{\mu_{t/2},\dots\mu_{t}\}$. At early times, when both $\mu_{t}$
and $N_{t}$ are varying significantly, it is assured that $\kappa_{t}^{\mathrm{max}}\sim\overline{\kappa}_{t}\sim V$.
At late times, $\mu_{t}$ should settle to the target chemical potential,
whereas $N_{t}$ will continue to exhibit equilibrium fluctuations.
Then $\kappa_{t}^{\mathrm{max}}$ grows very large, and we expect
to recover the fluctuation-based estimator, $\overline{\kappa}_{t}=\kappa_{t}^{\mathrm{fluc}}$.

\subsubsection{Method summary}
Pseudocode for the full $\mu$-tuning algorithm is listed in Algorithm~\ref{alg:tuning}. The mean and variance estimators can be updated in constant time using
the methods described in Appendix~\ref{sec:welford}. The user must
provide an initial guess $\mu_{t=0}$ for the chemical potential. Also required is a parameter $\alpha$ that sets an approximate scale for the compressibility.

We make two final remarks regarding the algorithm. Note, first, that
convergence, $\mu_{t+1}=\overline{\mu}_{t}$, implies that the target
condition is satisfied, $\overline{N}_{t}=N^{\ast}$. Second, the
dynamical update rules are inherently self-stabilizing. Suppose that
changes to $\mu_{t}$ are having no major effect on $N_{t}$. Then
the sample variance $\overline{\mathrm{Var}}_{t}[N]$ will decrease,
leading to smaller $\overline{\kappa}_{t}$. This, in turn, will drive
larger updates to the chemical potential. Eventually the magnitude
of these updates will be enough to produce the necessary changes in
$N_{t}$. Conversely, if changes to $\mu_{t}$ are having too large
of an effect on $N_{t}$, then compressibility estimator $\overline{\kappa}_{t}$
will also grow large, and this will dampen the magnitude of updates
to $\mu_{t+1}$. These self-stabilizing mechanisms share conceptual similarities to those in standard
PID controllers~\cite{RevModPhys.77.783, Kleiner19}, though all parameters in our algorithm are physically motivated and work robustly across a range of systems.

\begin{algorithm}
\KwIn{Target particle number $N^\ast$}
\KwIn{Initial guess $\mu_{t=0}$ for the chemical potential}
\KwIn{Characteristic compressibility scale $\alpha$}
\For{$t = 0,1,\,\hdots $}{
    Perform MC sampling with chemical potential $\mu_t$ \\
    Collect samples for $\langle N \rangle$ and (in the QMC context) for $\langle N^2 \rangle$ \\
    Update time averages $\overline{\mu}_t$ and $\overline{N}_t$ \\
    Update variance estimators $\overline{\mathrm{Var}}_t [\mu]$ and $\overline{\mathrm{Var}}_t [N]$ \\
	Calculate $\kappa$ estimate $\kappa^{\mathrm{fluc}}_t = \beta \overline{\mathrm{Var}}_t [N]$ \\
	Calculate lower bound $\kappa^{\min}_t = \alpha / (t+1)^{1/2}$ \\
    Calculate upper bound $\kappa^{\max}_t = \sqrt{ \overline{\mathrm{Var}}_t [N] / \overline{\mathrm{Var}}_t [\mu]}$ \\
    Calculate bounded $\kappa$ estimate $\overline{\kappa}_t = \max(\kappa^{\min}_t, \min(\kappa^{\max}_t, \kappa^{\mathrm{fluc}}_t))$ \\
	Update $\mu_{t+1} = \bar{\mu}_{t} + (N^\ast - \overline{N}_t) / \overline{\kappa}_t$
}

\caption{Chemical potential tuning. In the language of magnetism, we would make the substitutions of Eq.~\eqref{eq:mag_subs}.}
\label{alg:tuning}
\end{algorithm}

\section{Results\label{sec:Results}}

\begin{figure*}
\centering
\includegraphics[width=2\columnwidth]{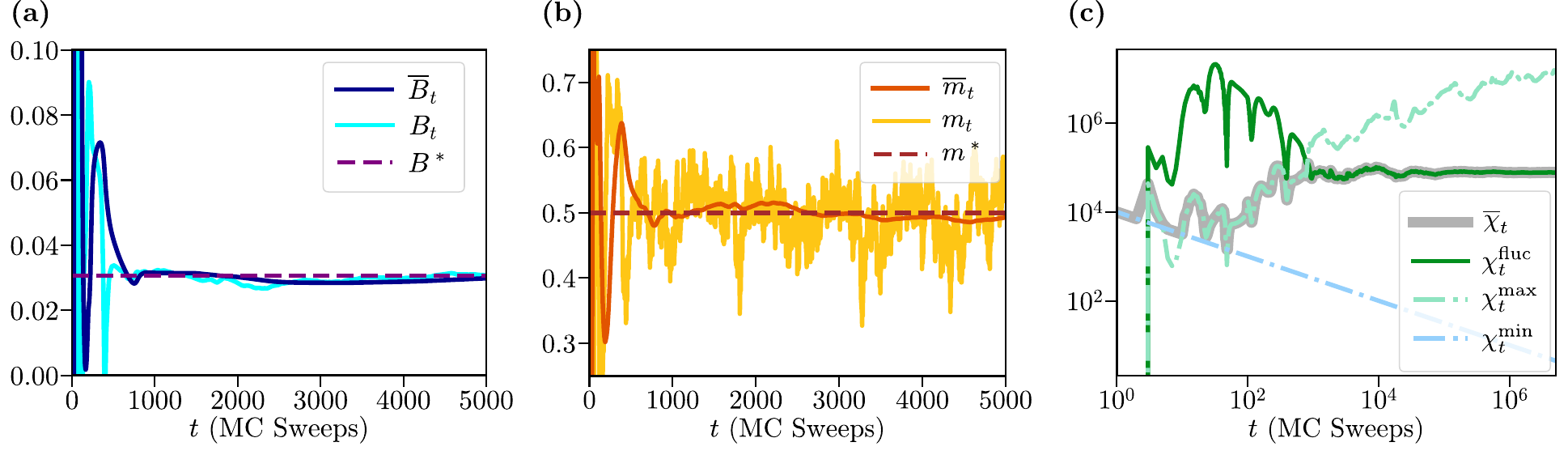}
\caption{Tuning the magnetic field $B$ of the classical Ising model to achieve a
target magnetization per site, $m^{\ast}=1/2$. We employ a $100\times100$ square lattice, and temperature $T=2.5$.
(a) The
dynamically evolving field $B_{t}$ and its running average $\overline{B}_{t}$
both eventually converge to  $B^{\ast}\approx3.096(2)\times10^{-2}$.
(b) Similar plots for $m_{t}=M_{t}/N_{\mathrm{sites}}$ and its running
average $\overline{m}_{t}$; significant equilibrium fluctuations
in $m_{t}$ are observed. (c) The susceptibility $\chi$ (analogous to $\kappa$) is reasonably estimated by  $\overline{\chi}_{t}$, with the lower and upper bounds, $\chi_{t}^{\mathrm{min}}$
and $\chi_{t}^{\mathrm{max}}$, playing important roles at early times. At long times the susceptibility converges to $7.76(4) \times 10^4$. Note that panel (c) is a log-log plot and extends over longer times than panels (a) and (b)}
\label{fig:ising_ex}
\end{figure*}

\subsection{Ising Model\label{subsec:Ising}}

We begin by demonstrating our method on a well-understood test case,
the ferromagnetic Ising model on a two-dimensional square lattice. The Hamiltonian is
\begin{equation}
H=-J\sum_{\langle ij\rangle}s_{i}s_{j}-B\sum_{i}s_{i},
\end{equation}
where the sum over $\langle ij\rangle$ is over all nearest neighbor
sites, and $s_{i}=\pm1$. For simplicity, all energies will be measured
in units of $J=1$ and we likewise set the Boltzmann constant
$k_{\rm B}=1$. The tunable quantity here is the magnetic
field $B$, which couples linearly to the total magnetization $M=\sum_{i}s_{i}$. In this case, the magnetic susceptibility $\chi=\d\langle M\rangle/\d B=\beta\left(\langle M^{2}\rangle-\langle M\rangle^{2}\right)$ plays the role of the compressibility $\kappa$. 
Algorithm~(\ref{alg:tuning}) applies upon making the following substitutions
\begin{equation}
\mu\mapsto B,\quad N\mapsto M,\quad \kappa \mapsto \chi.\label{eq:mag_subs}
\end{equation}
It will also be convenient to refer to the magnetization per site, $m=M/N_{\mathrm{sites}}$.

In zero field, this system undergoes a second-order phase transition
between the paramagnetic and ferromagnetic phases at a critical temperature
\textbf{$T_{c}\approx2.27$}~\citep{Onsager44}. Approaching $T_{c}$
from above causes both the susceptibility $\chi$ and the autocorrelation
time to diverge. The distance $T-T_{c}$ from the critical point offers
an excellent means to scale the ``tuning difficulty''. 

To begin, we investigate an example tuning run on a $100\times100$
Ising system at a temperature of $T=2.5$, starting with a uniformly
random initial state. By symmetry, zero magnetization is achieved
at zero field. To make the tuning task more interesting, we aim to
find the magnetic field $B=B^{\ast}$ that produces a \emph{nonzero
} target magnetization-per-site of $m^{\ast}=1/2$. Despite the presence of a small but finite field $B^\ast > 0$, we still observe very large autocorrelation times when $T$ approaches $T_c$. To explore this effective critical slowing down, we opt to use the standard single spin-flip Metropolis-Hastings
algorithm~\citep{Hastings70}, though more advanced cluster updates~\citep{Wang90, Kent-Dobias2018Phys.Rev.E}
would also be compatible with the tuning algorithm. The time
index $t$ is incremented once per MC sweep, at which point a new
measurement of the total magnetization $M_{t}$ is taken, and the
field $B_{t+1}$ is computed according to Algorithm~\ref{alg:tuning}
under the substitutions of Eq.~(\ref{eq:mag_subs}). \textcolor{black}{The
initial field is $B_{t=0}=0$.} To set a scale for $\chi_{t}^{\min}$,
we select $\alpha=N_{\mathrm{sites}}/J$.

\begin{figure}[b!]
\centering
\includegraphics[width=\columnwidth]{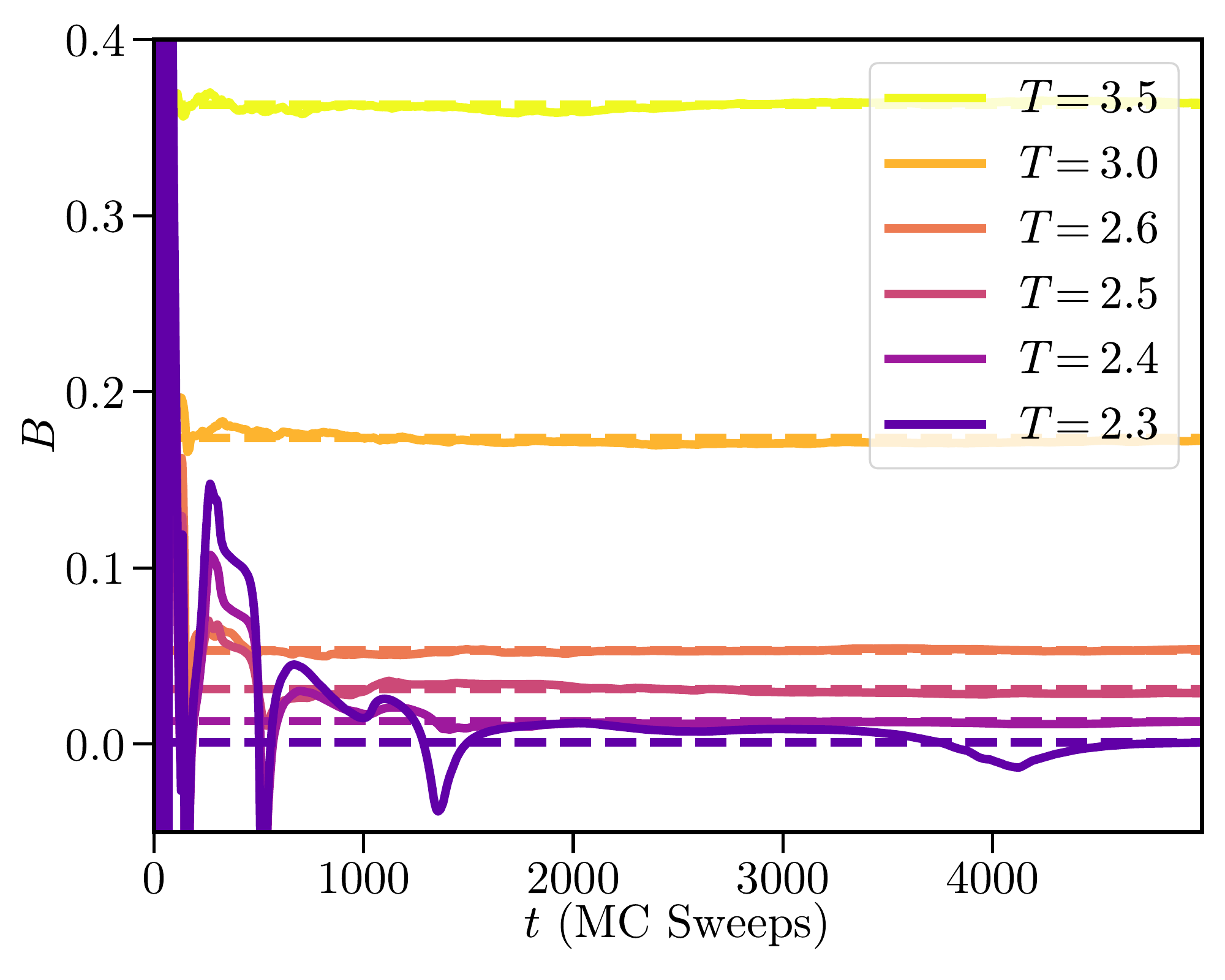}

\caption{Dynamically tuning the magnetic field $B_{t}$ (solid lines) towards the solution $B^{\ast}$ (dashed lines) which achieves a target
magnetization per site $m^{\ast}=1/2$ for various temperatures above
$T_{c}\approx2.27$. The most challenging case tested is $T=2.3$,
for which early-time values of $B_{t}$ can go as high as $2$}
\label{fig:ising_converge}
\end{figure}

In Fig.~\ref{fig:ising_ex}, we plot the result of dynamically tuning
the field $B_{t}$. Panels~\ref{fig:ising_ex}(a) and~\ref{fig:ising_ex}(b)
show the time evolution of $B_{t}$ and $m_{t}$, and their time
averages. Panel~\ref{fig:ising_ex}(c) shows estimators for the susceptibility $\chi$ (compressibility $\kappa$ in lattice gas language). The dynamics undergoes three different
regimes, corresponding to the three branches in the expression $\overline{\chi}_{t}=\max\left[\chi_{t}^{\mathrm{min}},\min\left(\chi_{t}^{\mathrm{max}},\chi_{t}^{\mathrm{fluc}}\right)\right]$.
For the first few MC sweeps ($t\leq3)$ there is essentially no good
susceptibility data. Here, the lower bound $\chi_{t}^{\mathrm{min}}$
of Eq.~(\ref{eq:kappa_min}) controls the estimator $\overline{\chi}_{t}$
and prevents the algorithm from making overly large corrections
to $B_{t}$. In the intermediate time regime of $3<t\lesssim10^{3}$,
the field $B_{t}$ is evolving significantly, and driving large changes
to $m_{t}$. Here, the upper bound $\chi_{t}^{\mathrm{max}}$ of
Eq.~(\ref{eq:kappa_max}) controls the estimator $\overline{\chi}_{t}$
and correctly captures the approximate sensitivity of magnetism to
changes in the applied field. By the end of this regime, fluctuations
in $B_{t}$ decrease significantly. Finally, at times $t\gtrsim10^{3}$,
the fluctuation-based estimator $\chi_{t}^{\mathrm{fluc}}$ becomes
accurate, and $B_{t}$ converges precisely toward the solution
$B^{\ast}\approx3.096(2)\times10^{-2}$.

\begin{figure*}[t]
\centering
\includegraphics[width=2\columnwidth]{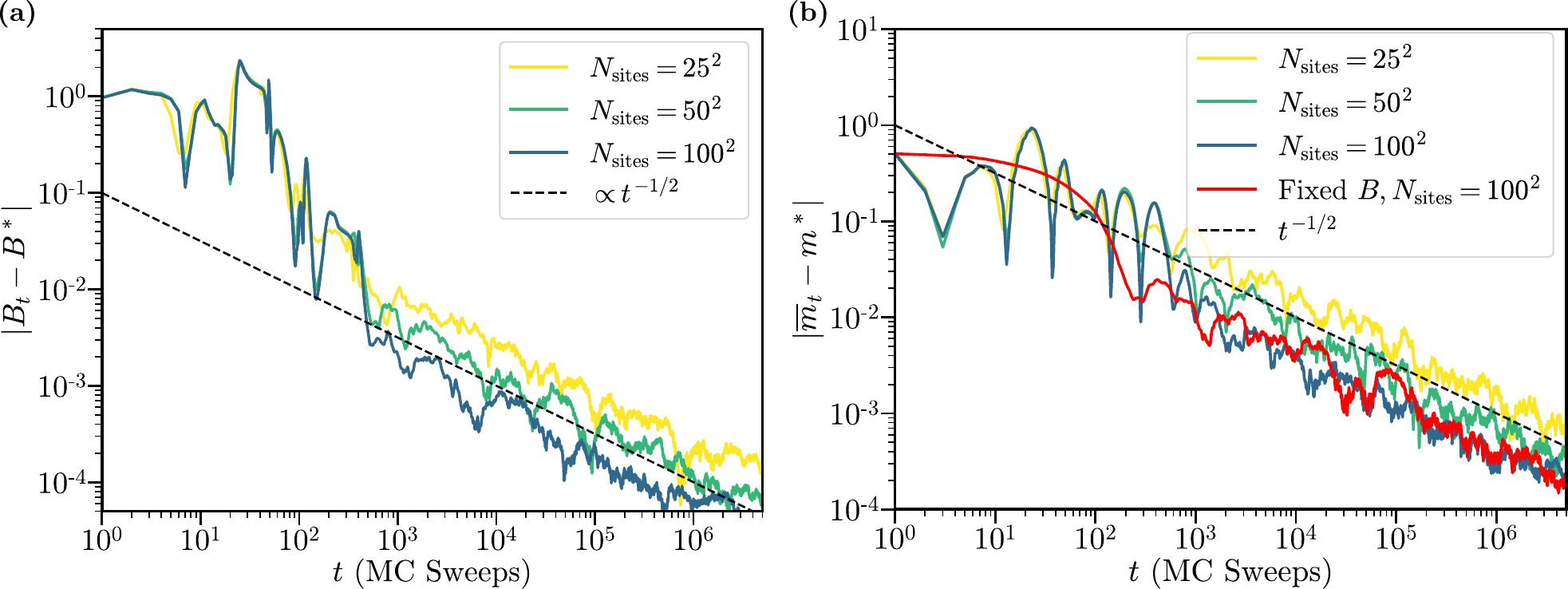}
\caption{Errors in (a) the dynamically-tuned magnetic field and
(b) the running-mean magnetization. We consider Ising systems of various sizes,
with $T=2.5$ and target magnetization $m^{\ast}=0.5$. Error curves are averaged over 10 independent runs. The reference magnetizations $B^{\ast}$ are estimated, per system size, by averaging $B_{t}$ at $t=5\times10^{6}$ over all 10 runs.
The red curve in (b) results from a $N_{\mathrm{sites}}=100^{2}$
simulation where $B = B^\ast$ is held fixed throughout
the entire simulation}
\label{fig:ising_error}
\end{figure*}

In Fig.~\ref{fig:ising_converge}, we show the evolution of $B_{t}$
for the same Ising model, but now over a variety of temperatures.
At high $T$, the tuner converges to the correct value $B^{\ast}$
very quickly. At temperatures approaching $T_{c}\approx2.27$, both
the autocorrelation time and the magnetic susceptibility grow large,
leading to difficulty in collecting accurate statistics, and very
high sensitivity to small changes in the applied field. Despite
the slower convergence near $T_{c}$, the tuner appears
to be working correctly in all cases tested.

Figure~\ref{fig:ising_error} shows the error of the instantaneous
field $|B_{t}-B^{\ast}|$ and the running-mean magnetization $|\bar{m}_{t}-m^{\ast}|$,
throughout the tuning process. Here we take $T=2.5$ and $m^{\ast}=1/2$
as in Fig.~\ref{fig:ising_ex}, and allow system size to vary. Our best estimates
for $B^{\ast}$ are obtained by averaging over 10 independent runs,
extending to $t=5\times10^{6}$ Monte Carlo sweeps. The results are $B^{\ast}=3.113(3)\times10^{-2}$ and $B^{\ast}=3.096(1)\times10^{-2}$, for lattice sizes $N_{\mathrm{sites}}=25^{2}$
and $100^{2}$, respectively. The system size dependence is relatively minor, given our choice of $T=2.5$. At temperatures nearer to $T_c \approx 2.27$, the $B^\ast$ would become smaller, and their relative variation would depend more strongly on system size.

The early time dynamics (up to $\sim10^{2}$ sweeps) of both errors
are seen to be system-size independent due to the dynamics being dominantly
controlled by our $\chi^{\mathrm{min}}$ and $\chi^{\mathrm{max}}$
bounding scheme. Once $\chi^\mathrm{fluc}$ gains control, we can see a
separation emerge as larger systems enjoy improved statistics from
increased self-averaging, resulting in faster tuning. At these large
times, the errors decay as $t^{-1/2}$, and the limiting factor in
tuning becomes the statistical error in the Monte Carlo estimates.
In this regime, note that the error in the average magnetization,
$|\bar{m}_{t}-m^{\ast}|$, is roughly independent of whether $B_{t}$
is being tuned dynamically, or fixed to the correct value $B=B^{\ast}$
throughout the simulation [shown as the red curve in Fig.~\ref{fig:ising_error}(b)].

\subsection{Holstein Model\label{subsec:Holstein}}

We now demonstrate our method in a quantum Monte Carlo (QMC) setting. As a test case we consider the Holstein model, one of the simplest models describing
interactions between electrons and phonons on a lattice \cite{Holstein59}. The Hamiltonian
is
\begin{align}
\hat{H} & =-t_{\mathrm{h}}\sum_{\langle\boldsymbol{i}\boldsymbol{j}\rangle,\sigma}(\hat{c}_{\boldsymbol{i},\sigma}^{\dagger}\hat{c}^{\phantom\dagger}_{\boldsymbol{j},\sigma}+\mathrm{h.c.}) -\mu\sum_{\boldsymbol{i},\sigma}\hat{n}_{\boldsymbol{i},\sigma} \nonumber \\
 & \quad+\frac{1}{2}\sum_{\boldsymbol{i}}\hat{P}_{\boldsymbol{i}}^{2}+\frac{\omega^{2}}{2}\sum_{\boldsymbol{i}}\hat{X}_{\boldsymbol{i}}^{2}+\lambda\sum_{\boldsymbol{i},\sigma}\hat{n}_{\boldsymbol{i},\sigma}\hat{X}_{\boldsymbol{i}}.
 \label{eq:holstein}
\end{align}
The operator $\hat{c}_{\boldsymbol{i},\sigma}^{\dagger}$ creates
an electron on site $\boldsymbol{i}$ with spin $\sigma$, and $\hat{n}_{\boldsymbol{i},\sigma}=\hat{c}_{\boldsymbol{i},\sigma}^{\dagger}\hat{c}^{\phantom\dagger}_{\boldsymbol{i},\sigma}$
is the electron number. The first term in $\hat{H}$ describes hopping
between nearest-neighbor sites $\left\langle \boldsymbol{i}\boldsymbol{j}\right\rangle $.
The Hamiltonian also includes bosonic position and momentum operators,
$\hat{X}_{\boldsymbol{i}}$ and $\hat{P}_{\boldsymbol{i}}$, which
models a local phonon mode on site $\boldsymbol{i}$ with frequency
$\omega$. The term proportional to $\lambda$ couples the electrons
and phonons leading to a phonon mediated effective electron-electron
interaction.
We set our energy units in terms of the hopping amplitude $t_{\mathrm{h}}=1$,
in which the value of the chemical potential needed to obtain half-filling (i.e. one electron per site on average) is known
to be $\mu=-\frac{\lambda^{2}}{\omega^{2}}$ from a particle-hole transformation~\cite{PhysRevB.87.235133}. Due to the Pauli exclusion principle, at most two electrons can exist on a single
site.

The total electron number operator is $\hat{N}=\sum_{\boldsymbol{i},\sigma}\hat{n}_{\boldsymbol{i},\sigma}.$
Our goal is to tune the chemical potential $\mu$ to produce a target
density, $\langle \hat{N}\rangle =N^{\ast}$. Here, the expectation
of an observable $\hat{\mathcal{O}}$ is understood to mean
\begin{align}
\left\langle \hat{\mathcal{O}}\right\rangle 
&= {\cal Z}^{-1}
\mathrm{Tr}\,e^{-\beta\hat{H}}\hat{\mathcal{O}}, \label{eq:observable}
\\
{\cal Z}&=
\mathrm{Tr}\,e^{-\beta\hat{H}
},
\end{align}
and the trace runs over the entire Fock space.

\subsubsection{Overview of quantum Monte Carlo}

There are many forms of QMC. One widely used set of methods
begins by expressing the
many-body partition function $Z=\mathrm{Tr}\,e^{-\beta\hat{H}}$ as
a path integral involving fields that fluctuate in imaginary time.
The aim is then to perform
\emph{ordinary} Monte Carlo sampling of these fluctuating variables
according to some appropriate probability distribution. In the determinant-QMC approach~\citep{Blankenbecler81}, which is our focus here, one uses a Suzuki-Trotter expansion inside the trace 
\begin{equation}
Z=\mathrm{Tr}\,\underset{\beta/\Delta t\textrm{ imaginary time slices}}{\underbrace{e^{-\Delta\tau\hat{H}_{0}}e^{-\Delta\tau\hat{H}_{1}}\dots e^{-\Delta\tau\hat{H}_{0}}e^{-\Delta\tau\hat{H}_{1}}}}+\mathcal{O}(\Delta\tau^{2}),
\end{equation}
with a carefully selected decomposition $\hat{H}=\hat{H}_{0}+\hat{H}_{1}$.
Inserting a complete set of states at each discrete imaginary time
slice, $0\leq\tau<\beta$, introduces an effectively classical field $x_{\tau,\boldsymbol{i}}$ which, when sampled, allows to 
estimate observables $\left\langle \hat{\mathcal{O}}\right\rangle $.
Each sample $x_{\tau,\boldsymbol{i}}$ is typically weighted according
to a fermionic determinant, $P[x]\propto\left|\det M_{\uparrow}^{\dagger}M_{\downarrow}\right|$,
for an appropriate matrix function $M_{\sigma}[x]$.%

When this procedure is applied to the Holstein model, the field $x_{\tau,\boldsymbol{i}}$
can be interpreted as ``imaginary time fluctuations'' of the phonons.
An analogous formalism is used for sampling the gluon field in lattice
quantum chromodynamics (QCD), and we can borrow techniques from that
community. In particular, Langevin~\citep{Batrouni85} and hybrid
Monte Carlo (HMC) sampling~\citep{Duane87,Neal11} have both proven
effective for simulating electron-phonon models~\citep{Batrouni19,Beyl18},
and make it possible to update the entire field $x_{\tau,\boldsymbol{i}}$
at a cost that scales near-linearly with system size. Here we employ
HMC. A complete account of our QMC methodology is presented in~\cite{Bensawesomepaper}.

\subsubsection{Chemical potential tuning for quantum models}

Algorithm~\ref{alg:tuning} remains valid in the QMC context provided that we are careful in estimating the compressibility $\kappa$. The thermodynamic relationship of Eq.~(\ref{eq:kappa}) continues to
hold,
\begin{equation}
\kappa=\frac{\d\langle\hat{N}\rangle}{\d\mu}=\beta\mathrm{Var}[\hat{N}],\label{eq:kappa-1}
\end{equation}
where $\mathrm{Var}[\hat{N}]=\langle\hat{N}^{2}\rangle-\langle\hat{N}\rangle^{2}$, and each expectation value on the right-hand side
is to be interpreted in the sense of Eq.~(\ref{eq:observable}). An interesting feature of QMC, however,
is that unbiased samples $N_{t}$ of $\langle \hat{N}\rangle$ do {\em not} generally contain sufficient information to estimate 
the variance of $\hat{N}$ due to neglecting within-sample fluctuations~\footnote{Classical microstates sampled from the grand canonical equilibrium distribution have a well defined particle number. Our QMC simulations, in contrast, sample a phonon field that fluctuates in imaginary time, and each phonon configuration is associated with an entire statistical distribution of electron numbers.
}.
We can still define a fluctuation-based estimator in the form of Eq.~(\ref{eq:kappa_fluc}),
\begin{equation}
\kappa_{t}^{\mathrm{fluc}}=\beta\overline{\mathrm{Var}}_{t}[\hat{N}],
\end{equation}
but now we must use
\begin{equation}
\overline{\mathrm{Var}}_{t}[\hat{N}]=\overline{N_{t}^{(2)}}-\left(\overline{N}_{t}\right)^{2},\label{eq:quantum_var}
\end{equation}
where $N_{t}^{(2)}$ denotes a statistical sample of the expectation value $\langle \hat{N}^{2}\rangle $ defined in Eq.~(\ref{eq:observable}).
Time-averages have the same form as in Eqs.~(\ref{eq:mu_bar}) and~(\ref{eq:N_bar}),
but now we must also track, 

\begin{equation}
\overline{N_{t}^{(2)}}=\frac{1}{L_{t}}\sum_{t^\prime=\left\lceil t/2\right\rceil }^{t}N_{t^\prime}^{(2)}.
\end{equation}
The quantity $\kappa_{t}^{\max}$ remains as in Eq.~(\ref{eq:kappa_max}),
but using the sample variance of Eq.~(\ref{eq:quantum_var}). With
these refinements to $\kappa_{t}^{\mathrm{fluc}}$ and $\kappa_{t}^{\max}$,
we can directly apply Algorithm~\ref{alg:tuning} to tune the chemical
potential.%

\subsubsection{Single-Site Limit\label{subsec:holstein1d}}

\begin{figure*}[!ht]
\centering
\includegraphics[width=2\columnwidth]{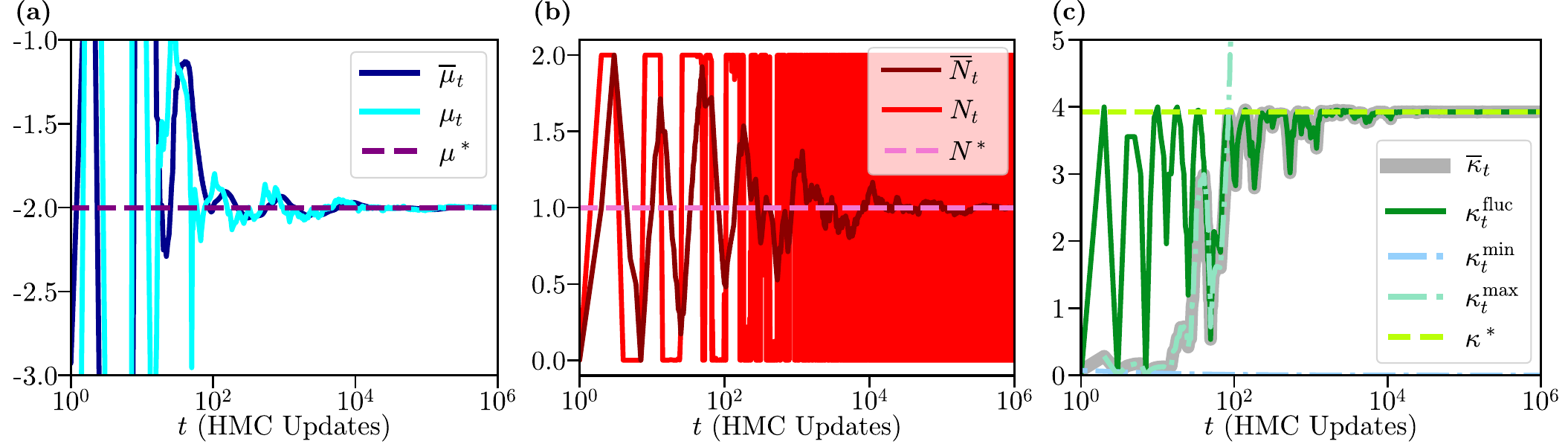}
\caption{Tuning the chemical potential \textbf{$\mu_{t}$} of a single-site
Holstein model with $\omega=1$, $\lambda=\sqrt{2}$, and $\beta=4$
to achieve half-filling, \textbf{$N^{*}=1$}. The target chemical
potential is exactly $\mu^{\ast}=-2$. Early time fluctuations in
$\mu_{t}$ are of order 10. In the absence of dynamical $\mu$-tuning,
the natural transition rate between metastable wells is of order $\Delta t=10^{2}$}

\label{fig:single_site}
\end{figure*}

As an initial demonstration, we examine the behavior of Algorithm~\ref{alg:tuning}
in the single-site limit of the Holstein model $\left(t_{\mathrm{h}}=0\right)$.
This model admits an analytic solution by the Lang-Firsov transformation \cite{lang1963zh, mah00}, but nonetheless serves as
a challenging test-case for $\mu$-tuning. The effective action $S[x]$ resulting from Eq.~(\ref{eq:holstein}) possesses two deep local minima (associated
with electron number 0 or 2) due to an effective electron-electron
attraction that is mediated by the phonons. MC sampling of the phonon
field $x_{\tau}$ is characterized by long periods of trapping within
one minima, punctuated by rare hops across the barrier into the other.
Strong hysteresis creates a challenge for $\mu$-tuning; adjusting
the chemical potential $\mu_{t}$ may affect measurements of electron
number $N_{t^\prime}$ only after a very large amount of simulation time,
$t^\prime \gg t$.

We consider a model with $\omega=1,\lambda=\sqrt{2}$ at $\beta=4$.
These are values commonly used in studies of the charge density
wave transition in the Holstein model, which occurs at 
$\beta_c \sim 6$ when $t_h = 1$ \cite{FengPRB2020}.
We aim for a target electron number of $\langle\hat{N}\rangle=1$,
i.e. half filling. This corresponds to finding the chemical potential
at which the average occupancies of the two metastable states are
equal. In the single site limit at half filling, the exact chemical
potential and compressibility are $\mu^{\ast}=-\lambda^{2}/\omega^{2}$
and $\kappa^{\ast}=\beta/(1+e^{\beta\mu^{*}/2})$. We use an initial
guess for the chemical potential, $\mu_{0}=0$, that is purposefully
distant from the true value $\mu^{\ast}=-2$. To set a scale for $\kappa_{t}^{\min}$,
we select $\alpha=N_{\mathrm{sites}}/\omega$. 

Figure~\ref{fig:single_site}
illustrates a representative $\mu$-tuning run. Each increment in
time $t$ corresponds to a single HMC trial update. At early times,
Fig.~\ref{fig:single_site}(b) shows sharp transitions (``jumps'')
in the measurements $N_{t}$, which are largely driven by changes
in $\mu_{t}$. These jumps correspond to transitions between the two
metastable wells, and in the absence of $\mu$-tuning, would occur
on the time-scale of $10^{2}$ HMC trial updates.

Figure~\ref{fig:single_site}(c) shows that each of the early-time
jumps between metastable wells is accompanied by a large spike in
$\bar{\kappa}_{t}$, which reflects the
large change in $N_{t}$. After each jump, there is a significant
period of time where $N_{t}$ is roughly constant, which causes $\bar{\kappa}_{t}$
to drop. At times $t\lesssim100$, the upper bound $\kappa^{\mathrm{max}}$
is instrumental in allowing the tuning dynamics to make significant
corrections to $\mu$, which drive the density back and forth between
metastable wells on an exponentially growing time scale.\textcolor{black}{{}
At around $t\approx100$, we reach the time scale required for
natural (equilibrium) jumps between the the two metastable wells.
At this point, the algorithm} switches over to the fluctuation-based
compressibility estimator $\bar{\kappa}_{t}=\kappa_{t}^{\mathrm{fluc}}$,
as observed Fig.~\ref{fig:single_site}(c). At times $t\gtrsim100$, the errors in statistical estimators decay like $t^{-1/2}$ in a controlled fashion.

\subsubsection{Full 2D system\label{subsec:holstein2d}}

Finally, to benchmark our algorithm in a more realistic setting, we
consider a square lattice Holstein model with phonon frequency $\omega=1$
and coupling strength $\lambda=\sqrt{2}$. We consider a square lattice
of size $L=10$, with $N_{{\rm sites}}=L^{2}$ total sites. At half-filling
the ground state is characterized by a finite temperature phase transition
to charge-density-wave (CDW) order, where the electrons localize onto
one of the two sublattices, spontaneously breaking a $\mathbb{Z}_{2}$
symmetry. For our chosen parameters the critical inverse temperature
is approximately $\beta\sim6$.

We test our algorithm at an inverse temperature $\beta=10$. At half-filling,
this low temperature gives rise to a gapped CDW phase. When doped
sufficiently away from half filling, and at sufficiently low temperature,
the system is expected to transition to a superconducting phase\cite{bradley21, PhysRevB.103.235156}.

We measure the CDW order using the staggered charge susceptibility
\begin{align}
\chi_{{\rm cdw}}=\int_{0}^{\beta} & \sum_{\boldsymbol{r}}\left(-1\right)^{(r_x+r_y)}C\left(\boldsymbol{r},\tau\right)d\tau,\label{eq:Scdw}
\end{align}
defined in terms of the real-space density-density correlation function
\begin{align}
C\left(\boldsymbol{r},\tau \right)= & \frac{1}{N}\sum_{\boldsymbol{i}}\left\langle 
\hat n_{\boldsymbol i + \boldsymbol r}(\tau) \hat n_{\boldsymbol i}(0) 
\right\rangle,\label{eq:DenDen}
\end{align}
where $\hat n_{\boldsymbol i}(\tau) = \hat n_{\boldsymbol i,\downarrow}(\tau) + \hat n_{\boldsymbol i,\uparrow}(\tau)$ denotes the total electron number on site $\boldsymbol i$ at imaginary time $\tau$.
A signature for superconducting order is given by the pair susceptibility
\begin{align}
P_{s}= & \frac{1}{N_{{\rm sites}}}\int_{0}^{\beta}\left\langle \hat{\Delta}\left(\tau\right)\hat{\Delta}^{\dagger}\left(0\right)\right\rangle d\tau,
\end{align}
where $\hat{\Delta}\left(\tau\right)=\sum_{i}\hat{c}_{\boldsymbol i,\downarrow}\left(\tau\right)\hat{c}_{\boldsymbol i,\uparrow}\left(\tau\right)$.

\begin{figure}
\includegraphics[width=0.95\columnwidth]{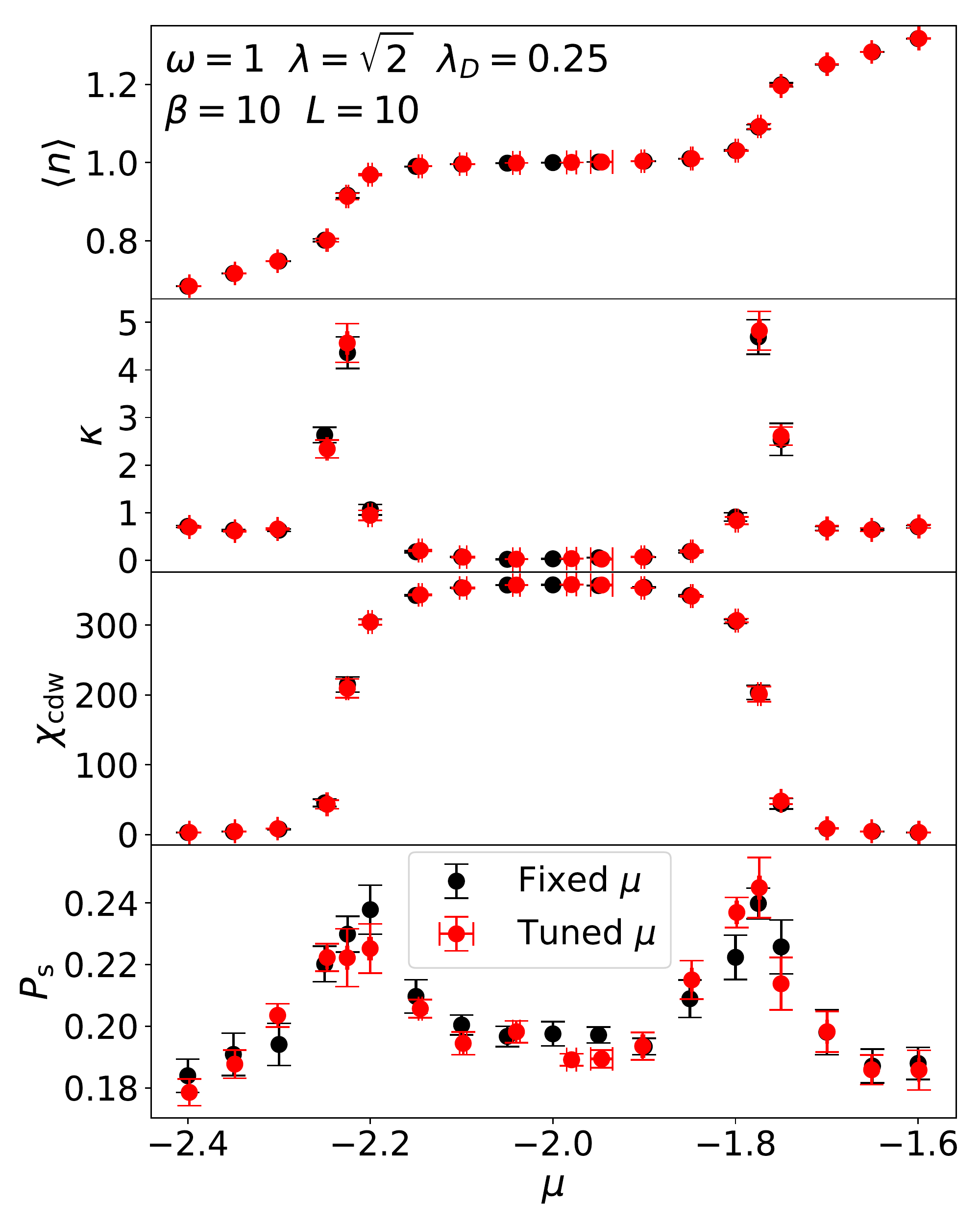}
\caption{Comparison of several observables measured
in a $10\times10$ Holstein model under the fixed-$\mu$ scheme (black),
and under dynamic $\mu$-tuning scheme aiming to achieve a target
density (red). Horizontal error bars on the red markers
indicate uncertainty as measured by the standard error of $\mu$}
\label{fig:holstein_cdw}
\end{figure}

In Fig.~\ref{fig:holstein_cdw} we compare two sets of simulation
results, one where $\mu$ is held fixed, and the other where we tune
$\mu$ to a target density that was measured in the first set of results.
All simulations employed $t=5\times 10^3$ thermalization HMC steps.
Following this, we performed $5\times 10^4$ steps, with measurements
taken at every step. For the dynamical $\mu$ simulations we initialized
the chemical potential to $\mu_{t=0}=0$ and set $\alpha=N_{{\rm sites}}/\omega$.
The chemical potential was continually updated throughout the simulation,
but had largely converged already by the end of the thermalization
process. There is very good agreement between the two sets of data;
the error bars with $\mu$-tuning enabled are not discernibly larger
than with $\mu$ fixed to its target value.

Figure\ \ref{fig:holstein_cdw}(a) shows the density as a function
of the chemical potential. The plateau at half filling ($\mu=-2$)
illustrates the gapped CDW phase. The relatively large horizontal
error bars in the tuned value of $\mu$ near half-filling are associated
with a vanishing compressibility $\kappa$, Fig.\ \ref{fig:holstein_cdw}(b).
In other words, a fairly wide range of chemical potentials give rise
to (approximately) half-filling. Observe that the plateau in $\left\langle n\right\rangle = \langle N \rangle / N_{\mathrm{sites}}$
corresponds to a strong enhancement of $\chi_{\mathrm{cdw}}$ in Fig.\ \ref{fig:holstein_cdw}(c)
and a suppression of $P_{s}$ in Fig.\ \ref{fig:holstein_cdw}(d).
With enough doping, at approximately $\mu=-2.3$ and $\mu=-1.7,$
the marker $\chi_{\mathrm{cdw}}$ for CDW order rapidly vanishes.
Simultaneously, the density rapidly shifts away from half-filling,
as reflected by the two peaks in the compressibility $\kappa$. We
emphasize that the $\mu$-tuning algorithm performs well throughout
the diverse range of behaviors exhibited in this model.

\section{Discussion}

The dynamical $\mu$-tuning algorithm presented in Algorithm~\ref{alg:tuning}
enables simulation in the grand-canonical ensemble while targeting
a fixed mean particle number. The algorithm is straightforward to
implement, and imposes negligible computational overhead. Note that all running
time averages can be updated in constant time using the formulas of Appendix~\ref{sec:welford}.

Under the proposed scheme, the chemical potential $\mu$ is adjusted concurrently with the MC sampling. Although we do not provide formal convergence guarantees, the method works well in practice. For long-running simulations, most statistical samples will be collected after $\mu$ has approximately converged to its target value, and our benchmarks show that errors are well controlled.

Models with long autocorrelation times present a practical challenge,
in that it becomes difficult to assess the impact of a modified
$\mu$ value on the resulting mean particle number $\left\langle N\right\rangle $.
Our solution is to effectively collect statistics
over increasingly large time-windows such that, eventually, both $\left\langle N\right\rangle $
and its sensitivity $\kappa=\mathrm{d}\left\langle N\right\rangle /\mathrm{d}\mu$
can be accurately measured. We demonstrated
that our scheme works well even in very challenging cases, such as
the Ising model approaching criticality, and the single-site Holstein model
with strong metastability due to phonon-mediated electron binding
and associated large energy barriers.
The method also works well for larger-size quantum Monte Carlo simulations of the Holstein model on the square lattice, including at filling fractions coinciding with a charge density wave gap, where $\kappa$ approximately
vanishes.

Many variations of the $\mu$-tuning algorithm could be studied. For
example, one could modify the definition of the running time averages
in Eqs.~(\ref{eq:mu_bar}) and~(\ref{eq:N_bar}) to ``smoothly
forget'' past data, with the goal of reducing underdamped oscillations
in the early-time dynamics of the tuned field,
e.g., in Figs.~\ref{fig:ising_ex}(a) and~\ref{fig:single_site}(a). One might also explore whether ideas for accelerating fixed point solvers (e.g., Anderson mixing) could somehow be incorporated into Eq.~(\ref{eq:tuner}), which updates $\mu$ as a correction to the time-average $\overline \mu$. In our preliminary tests, however, we could not find any modifications to the algorithm that significantly improved accuracy over large simulation times.
Indeed, Figs.~\ref{fig:ising_error} and~\ref{fig:holstein_cdw}
suggest that the $\mu$-tuning algorithm is already close to optimal;
errors in long-time statistical measurements are observed to be about
the same, whether $\mu$ is dynamically tuned or statically fixed
to the exactly correct value.

\acknowledgments
K.~B. acknowledges support from the Center of Materials Theory as a part of the Computational Materials Science (CMS) program, funded by the U.S. Department of Energy, Office of Basic Energy Sciences. S.~J., O.~B., and R.~T.~S.~acknowledge support from the U.S. Department of Energy, Office of Science, Office of Basic Energy Sciences, under Award Number DE-SC0022311. C.~M. acknowledges support by the U.S. Department of Energy, Office of Science, Office of Advanced Scientific Computing Research, Department of Energy Computational Graduate Fellowship under Award Number DE-SC0020347. B.~C.-S.~was funded by a U.C.~National Laboratory In-Residence Graduate Fellowship through the U.C.~National Laboratory Fees Research Program. 

This report was prepared as an account of work sponsored by an agency of the United States Government.  Neither the United States Government nor any agency thereof, nor any of their employees, makes any warranty, express or implied, or assumes any legal liability or responsibility for the accuracy, completeness, or usefulness of any information, apparatus, product, or process disclosed, or represents that its use would not infringe privately owned rights.  Reference herein to any specific commercial product, process, or service by trade name, trademark, manufacturer, or otherwise does not necessarily constitute or imply its endorsement, recommendation, or favoring by the United States Government or any agency thereof.  The views and opinions of authors expressed herein do not necessarily state or reflect those of the United States Government or any agency thereof.

\section*{Author Declarations}

The authors have no conflicts to disclose.

\section*{Code Availability}

Code and an interactive interface to explore tuning for the Ising system is available at \url{https://colemiles.github.io/ising-tuner}. Code for
Holstein simulations, with $\mu$-tuning built-in, is available at \url{https://github.com/cohensbw/ElPhDynamics}. A general purpose implementation of the algorithm is provided by the Julia package \texttt{MuTuner.jl}, available at \url{https://github.com/cohensbw/MuTuner.jl}.

\appendix

\section{Updating running time-averages\label{sec:welford}}

The $\mu$-tuning algorithm require to maintain running averages of the
form
\begin{equation}
\overline{x}_{t}=\frac{1}{L_{t}}\sum_{t^\prime=\left\lceil ct\right\rceil }^{t}x_{t^\prime},\label{eq:x_bar}
\end{equation}
where $L_{t}=t-\left\lceil ct\right\rceil +1$ is the count of samples in the average, and $\left\lceil \cdot\right\rceil $
denotes the ceiling function. We selected $c=1/2$ for our study,
but other values $0 < c < 1$ are possible. The $x_{t}$ data could
be one of the following: the instantaneous chemical potential $\mu_{t}$,
a sample $N_{t}$ for the particle number, or (in the QMC context)
a sample $N_{t}^{(2)}$ for the particle number squared.

After each MC time step, we wish to update the running average from $\overline{x}_{t}$ to $\overline{x}_{t+1}$.
It is helpful to distinguish between two cases,
\begin{align*}
\textrm{Case A} & :\,\,\left\lceil c(t+1)\right\rceil =\left\lceil ct\right\rceil \\
\textrm{Case B} & :\,\,\left\lceil c(t+1)\right\rceil =\left\lceil ct\right\rceil +1.
\end{align*}
These are the only two possibilities given our assumptions that $0<c<1$
and $t$ is integer. 

In Case A we must add the new datapoint $x_{t+1}$ to the running
average. In Case B we must additionally remove the datapoint $x_{\left\lceil ct\right\rceil }$
from the running average. The update rule is then,
\begin{align}
\overline{x}_{t+1} & =\begin{cases}
(L_{t}\overline{x}_{t}+x_{t+1})/L_{t+1} & \textrm{Case A}\\
(L_{t}\overline{x}_{t}+x_{t+1}-x_{\left\lceil ct\right\rceil })/L_{t+1} & \textrm{Case B}.
\end{cases}
\end{align}
In Case A the number of datapoints increases by one, $L_{t+1}=L_{t}+1$,
whereas in Case B, $L_{t+1}=L_{t}$. We can therefore rearrange as,
\begin{equation}
\overline{x}_{t+1}=\begin{cases}
\overline{x}_{t}+(x_{t+1}-\overline{x}_{t})/L_{t+1} & \textrm{Case A}\\
\overline{x}_{t}+(x_{t+1}-x_{\left\lceil ct\right\rceil })/L_{t+1} & \textrm{Case B},
\end{cases}\label{eq:xbar_running}
\end{equation}
which improves numerical accuracy.

We are also interested in keeping a running estimate of the sample
variance,
\begin{equation}
\overline{\mathrm{Var}}_{t}[x]=\overline{x_{t}^{2}}-\left(\overline{x}_{t}\right)^{2},\label{eq:var_bad}
\end{equation}
or, for quantum observables, the closely related Eq.~(\ref{eq:quantum_var}).
The formula suggests that we maintain a running average $\overline{x_{t}^{2}}$,
through which the variance follows immediately. To estimate $\mathrm{Var}[\hat{N}]$
in QMC simulations, such a strategy may be necessary. Whenever possible,
however, direct numerical implementation of Eq.~(\ref{eq:var_bad})
should be avoided due to potentially large floating point round-off
error. A much improved algorithm was proposed by Welford~\citep{Welford62},
which we here adapt.

The sample variance can be equivalently written,
\begin{equation}
\overline{\mathrm{Var}}_{t}[x]=M_{t}/L_{t},
\end{equation}
where
\begin{equation}
M_{t}=\sum_{t^\prime=\left\lceil ct\right\rceil }^{t}\left(x_{t^\prime}-\overline{x}_{t}\right)^{2}.
\end{equation}

\begin{widetext}After a somewhat lengthy derivation, one finds the
recursion relation,

\begin{equation}
M_{t+1}=\begin{cases}
M_{t}+(x_{t+1}-\overline{x}_{t})(x_{t+1}-\overline{x}_{t+1}) & \textrm{Case A}\\
M_{t}+\left(x_{t+1}-x_{\left\lceil ct\right\rceil }\right)\left(x_{t+1}-\overline{x}_{t+1}+x_{\left\lceil ct\right\rceil }-\overline{x}_{t}\right) & \textrm{Case B},
\end{cases}
\end{equation}
which is numerically stable and easy to implement given
that we are already maintaining the running average $\overline{x}_{t}$.
Note that Cases A and B coincide when $x_{\left\lceil ct\right\rceil }=\overline{x}_{t}$.

\end{widetext}

\bibliography{mu_tuning}

\end{document}